\documentclass[dvips]{article}%

\usepackage{macros}
\usepackage{comment}
\allowdisplaybreaks[4]

\usepackage{times}

\newcommand{\zp}{\text{\o}}

\newcommand{\val}{\mathcal{V}}
\newcommand{\vals}{\mathcal{V}_\sigma}
\newcommand{\A}{\mathcal{A}}
\newcommand{\Ab}{\mathcal{A}_\bot}
\newcommand{\sopt}{I_\sigma}
\newcommand{\attr}{\textrm{Attr}}

\newcommand{\src}{\textrm{src}}

\title{Strategy Iteration using Non-Deterministic Strategies for Solving Parity Games}

\author{Michael Luttenberger\thanks{Institut f{\"u}r Informatik, Technische Universit{\"a}t M{\"u}nchen}\\ \texttt{luttenbe@model.in.tum.de}}

\begin{document}
\maketitle
\begin{abstract}
This article extends the idea of solving parity games
by strategy iteration to non-deterministic strategies: 
In a non-deterministic strategy a player restricts himself to some non-empty subset 
of possible actions at a given node, instead of limiting himself to exactly one action.

We show that a strategy-improvement algorithm by by Bj{\"o}rklund, Sandberg, and Vorobyov \cite{BSV04}
can easily be adapted to the more general setting of non-deterministic strategies.
Further, we show that applying the heuristic of ``all profitable switches'' (cf. \cite{BSV02})
 leads to choosing a ``locally optimal'' successor strategy
in the setting of non-deterministic strategies, thereby obtaining an easy proof of an algorithm by Schewe \cite{S07b}.

In contrast to \cite{BSV04}, we present our algorithm directly 
for parity games which allows us to compare it to the algorithm
by Jurdzinski and V{\"o}ge \cite{JV00}:
We show that the valuations used in both algorithm
coincide on parity game arenas in which one player can ``surrender''.
Thus, our algorithm can also be seen as a generalization
of the one by Jurdzinski and V{\"o}ge to non-deterministic strategies.

Finally, using non-deterministic strategies allows us to 
show that the number of improvement steps is bound from above by $O(1.724^{n})$.
For strategy-improvement algorithms, 
this bound was previously only known to be attainable by using randomization (cf. \cite{BSV02}).
\end{abstract}
\section{Introduction}
A parity game arena consists of a directed graph $G = (V,E)$
where every vertex belongs to exactly one of two players, 
called player $0$ and player $1$. 
Every vertex is colored by some natural number in $\{0,1,\ldots,d-1\}$.
Starting from some initial vertex $v_0$, a play of both players is an infinite
path in $G$ where the owner of the current node determines the next vertex.
In a parity game, the winner of such an infinite play is then defined by the parity
of the maximal color which appears infinitely often along the given play.

As shown by Mostowski \cite{Mos91}, and independently by Emerson and Jutla \cite{EJ91}, 
there exists a partition of $V$ in two sets
$W_0$ and $W_1$ such that player $i$ has a memoryless strategy, i.e.
a map $\sigma_i : V_i \to V$ which maps every vertex $v$ controlled by player $i$
to some successor $v$, so that player $i$ wins any play starting from some $w\in W_i$
by using $\sigma_i$ to determine his moves.

Interest in parity games arises as determining the winning set $W_0$ is
equivalent to deciding whether a given $\mu$-calculus formula holds w.r.t. to a given
Kripke structure, i.e. determining $W_0$ is equivalent to the model checking problem of $\mu$-calculus.
Further interest is sparked as it is known that solving parity games is in UP$\cap$co-UP \cite{Jur98:IPL},
but no polynomial time algorithm has been found yet. 

In this article we consider an approach for calculating the winning sets which is known as strategy iteration or strategy improvement,
and can be described as follows in the setting of games:
In a first step, a way for valuating the strategies of player $0$ is fixed, thereby
inducing a partial order on the strategies of player $0$.
Then, one chooses an initial strategy $\sigma : V_0 \to V$ for player $0$.
Iteratively (i) the current strategy is valuated, (ii) by means of this valuation
possible improvements of the current strategy are determined, i.e. pairs $(u,v)$ such
that $\sigma[u\mapsto v]$ is a strategy having a better valuation than $\sigma$, 
(iii) a subset of the possible improvements is selected and implemented yielding
a better strategy $\sigma': V_0 \to V$. 
These steps are repeated until no improvements can be found anymore.

Although this approach usually (using no randomization \cite{BSV02}) allows only to give a bound exponential in $\abs{V_0}$ on the number of iterations needed till termination, there is no family of games known 
for which this approach leads to a super-polynomial number of improvement steps.
It is thus also used in practice e.g. in compilers \cite{GS07:CSL}.

In particular, this approach has been successfully applied in several different scenarios like Markov decision
processes \cite{How60}, stochastic games \cite{HK66},
or discounted payoff games \cite{Puri}. 
Using reductions, these algorithms can also be used for solving parity games.
In 2000 Jurdzinski and V{\"o}ge \cite{JV00} presented the first strategy-improvement algorithm
for parity games which directly works on the given parity game without requiring
any reductions to some intermediate representation.
Although the algorithm by Jurdzinski and V{\"o}ge did not lead to a better upper bound on the 
complexity of deciding the winner of a parity game with $n$ nodes and $d$ colors (the algorithm in \cite{JV00} has
a complexity of $O( (n/d)^d)$ whereas the upper bound of $O( (n/d)^{d/2} )$ was already known at that time \cite{Jur00:STACS}), it sparked a lot of interest as the strategy-improvement process w.r.t. parity
games is directly observable and not obfuscated by some reduction.

In this article, we extend strategy iteration to {\em non-deterministic strategies}:
In a non-deterministic strategy a player is not required to fix a single successor for any vertex controlled by him 
instead he restricts himself to some non-empty subset of all possible successors.
Using non-deterministic strategies seems to be more natural, as it allows a player to only ``disable'' those moves along which the valuation of the current strategy decreases.
Our algorithm is an extension of an algorithm by Bj{\"o}rklund, Sandberg, and Vorobyov \cite{BSV04}
proposed in 2004. In particular, we borrow their idea of giving one of the two players the option to give up and ``escape'' an infinite play he would lose by introducing a sink.
In contrast to the original algorithm in \cite{BSV04} we present this extended algorithm directly for parity games
in order to be able to compare this algorithm directly with the one by Jurdzinski and V{\"o}ge,
and also in the hope that this might lead to better insights regarding the strategy improvement process.

Strategy iteration, as described above, chooses in step (iii) some subset of 
possible changes in order to obtain the next (deterministic) strategy. 
A natural question is how to choose 
this set of changes. Obviously, one would like to choose these sets in such a way
that the total number of improvements steps is as small as possible -- we call
this ``globally optimal''. As no efficient algorithm for determining these
sets is known, usually heuristics are used instead.
One heuristic applied quite often in the case of a binary arena is called ``all profitable switches'' \cite{BSV02}:
In a binary arena, given a strategy $\sigma: V_0 \to V$ we can refer to the successors of $v\in V_0$ by
$\sigma(v)$ and $\overline{\sigma(v)}$. A strategy improvement step then amounts to deciding for every node $v\in V_0$ whether 
to switch from $\sigma(v)$ to $\overline{\sigma(v)}$, or not. 
``All possible switches'' refers then
to the heuristic of switching to $\overline{\sigma(v)}$ of every $v\in V$ 
if this switch is an improvement w.r.t. the used valuation.
Transferring this heuristic to the setting of non-deterministic strategies
the heuristic becomes simply to choose the set of all possible improvements of the given strategy
as the new strategy considered in the next step.
We show that this simple heuristic leads to the ``locally optimal'' improvement,
i.e. the strategy which is at least as good as any other strategy
obtainable by implementing a subset of the possible improvements.
By applying this heuristic in every step we obtain a new, 
in our opinion more natural and accessible,  presentation of 
the algorithm by Schewe proposed in \cite{S07b}: There only valuations (referred to as ``estimations'' there),
and deterministic strategies are considered, whereas the strategy improvement process itself, and the connection to \cite{BSV04} are obfuscated.
Further, the algorithm in \cite{S07b} does not work directly on parity games,
and requires some unnecessary restrictions on the graph structure of the arena, e.g. only bipartite arenas are considered.

We then compare our algorithm using non-deterministic strategies to the one by Jurdzinski and V{\"o}ge \cite{JV00}.
This is not possible w.r.t. the algorithm in \cite{BSV04} or \cite{S07b} as these do 
not work directly on parity games.
Here, we can show that the valuation used in our algorithm, resp. in \cite{JV00}
coincide,
which readily allows us to conclude that the locally optimal
improvement obtained by our algorithm is always at least as good as 
any locally improvement obtainable by \cite{JV00}.%

We obtain an upper bound of $O( \abs{V}^2 \cdot \abs{E} \cdot ( \frac{\abs{V}}{d} + 1 )^{d} )$ for our algorithm
which is the same as the one obtainable when using deterministic strategies \cite{BSV04}.
So using non-deterministic strategies comes ``for free''.
Of course, w.r.t. to the sub-exponential bound of $\abs{V}^{O(\sqrt{\abs{V}})}$ obtainable for the algorithm by Jurdzinski, Paterson and Zwick \cite{JPZ06}, our algorithm is not competitive.
Still, we think that our algorithm is interesting as  strategy-iteration in practice only requires a polynomial number of improvement steps in general, as already mentioned above. 
In particular, we can show that the number of improvement steps done by our algorithm when using
the ``all profitable switches''-heuristic, and thus by the one by Schewe \cite{S07b}, is bounded by $O(1.724^{\abs{V_0}})$, 
whereas the best known upper bound for strategy iteration when using only deterministic strategies and no randomization
in the improvement selection is $O({2^{\abs{V_0}}}/{\abs{V_0}})$ \cite{BSV02}. In particular, the bound of $O(1.724^{\abs{V_0}})$ was previously known to be obtainable only be choosing the improvements randomly \cite{BSV02}.

\paragraph{Organization:} Section~2 summarizes the standard definitions and results regarding parity games.
In Section~3 we extend parity games by allowing player $0$ to terminate infinite plays in order 
to escape an infinite play he would lose. This idea was first stated in \cite{BSV04}.
We combine this with a generalization of the path profiles used in \cite{JV00} in order 
to get an algorithm working directly on parity games.
Section~4 summarizes our strategy improvement algorithm using non-deterministic strategies.
Section~5 then compares the algorithm presented in this article with the one by Jurdzinski and V{\"o}ge.

\section{Preliminaries}

In this section we repeat the standard definitions and notations regarding parity games.

An {\em arena $\A$} is given by $(V,E,o)$, if $(V,E)$ is a finite, directed graph, 
where $o : V \to \{0,1\}$ assigns each node an owner.
We denote by $V_i := o^{-1}(i)$ the set of all nodes belonging to player $i\in\{0,1\}$,
and write $E_i$ for $E \cap V_i \times V$.
Given some subset $V' \subseteq V$ we write $\A|_{V'}$ for the restriction of the arena $\A$
to the nodes $V'$.
A play $\pi \in V^\N \cup V^\ast$ in $\A$ is any maximal path in $\A$ where we assume
that player $i$ determines the move $(\pi(i),\pi(i+1))$, if $\pi(i) \in V_i$.
For $(V,E)$ a directed graph, and $s\in V$ a node we write $sE$ for the set of successors of $s$.

For $\A = (V,E,o)$ an arena, a (memoryless) {\em strategy of player $i$ (short: $i$-strategy)} ($i\in\{0,1\}$) is 
any subset $\sigma \subseteq E_i$ satisfying
$\forall s\in V_i : \abs{sE} >0  \Rightarrow \abs{s\sigma} > 0$,
i.e. a strategy does not introduce any new dead ends.
$\sigma$ is {\em deterministic}, if $\abs{s\sigma} \le 1$ for all $s\in V_i$.
We write $E_\sigma$ for $E_{\sigma} = E_{1-i} \cup \sigma$,
and $\A|_{\sigma}$ for $(V,E_\sigma,o)$.

We assume that the reader is familiar with the concept of attractors. For convenience, 
a definition can be found in the appendix.

A {\em parity game arena} $\mathcal{A}$ is given by $( V, E, o, c )$ where
$(V,E,o)$ is an arena with $vE\neq\emptyset$ for all $v\in V$, and $c: V \to \{0,1,\ldots,d-1\}$
 assigns each node a color.
The winner of a play $\pi$ in a parity game arena is given by
$\limsup_{i \in \N} c( \pi(i) ) \pmod 2.$
Given a node $s$, a strategy $\sigma \in E_i$ is a winning strategy for $s$ of player $i$,
if he wins any play in $\A|_{\sigma}$ starting from $s$.
Player $i$ wins a node $s$, if he has a winning strategy for it.
$W_i$ denotes the set of nodes won by player $i$.
As we assume that every node has at least one successor, there are only infinite plays in a parity game arena.
Wlog., we further assume that $c^{-1}(k) \neq \emptyset$ for all $k\in\{0,1,\ldots,d-1\}$ as we may otherwise
reduce $d$.
A cycle $s_0 s_1 \ldots s_{n-1}$ (with $s_{i+1 \pmod n}\in s_i E$)  
in a parity game arena $\mathcal{A}$ is called {\em $i$-dominated}, 
if the parity of its highest color is $i$.
Player $i$ wins the node $s$ using strategy $\sigma \subseteq V_i\times V$,
iff every cycle reachable from $s$ in $\mathcal{A}|_{\sigma}$ is $i$-dominated.

\begin{theorem}{\cite{Mos91,EJ91}}
For any a parity game arena $\A$ we have $W_0 \cup W_1 = V$. 
Player $i$ possesses a deterministic strategy $\sigma^\ast_i : V_i \to V$ 
with which he win every node $s\in W_i$.
\end{theorem}

\section{Escape Arenas}\label{sec:escape-arena}
In this section we extend parity games by allowing player $0$ to {\em escape} an
infinite play which he would loose w.r.t. the parity game winning condition:

Let $\A = (V,E,o)$ be a parity game arena. We obtain the arena $\Ab = (V_\bot,E_\bot,o_\bot)$ 
from $\A$  by introducing a sink $\bot$ $V_\bot := V \uplus \{\bot\}$
where only player $0$ can choose to play to $\bot$ ($E_\bot := E \cup V_0 \times \{\bot\}$).
The sink $\bot$ itself has no out-going edges, and we assume that player $0$ controls $\bot$ ($o_\bot :=o\cup \{(\bot,0)\}$ although this is of no real importance.
Although, this construction was first proposed in \cite{BSV04} we refer to $\Ab$ 
as {\em escape arena} in the style of \cite{S07b}.
As $\Ab$ itself is no parity game arena anymore, we have to define the winner of such a finite play as well.
For this we extend the definition of {\em color profile}, 
which was first stated in \cite{BSV03}, to finite plays:

For a given escape arena $\Ab$ using $d$ colors $\{0,1,\ldots,d-1\}$, 
we define the set $\mathcal{P}$ of {\em color profiles} by $\mathcal{P} := \Z^d\cup\{-\infty,\infty\}$ where $\Z^d$ is the set of $d$-dimensional integer vectors. 
We write $\zp$ for the zero-profile $(0,0,\ldots,0)\in \Z^d$,
and use standard addition on $\Z^d$ for two profiles $\wp,\wp'\in \Z^d$.
The idea of a profile $\wp \in \mathcal{P}$ is to count how often a given color
appears a long a finite play, whereas $-\infty$, reps. $\infty$ correspond
to infinite plays won by player $1$, resp. player $0$. 
More precisely, for a finite sequence $\pi = s_0 s_1 \ldots s_l$ of vertices, 
 the {\em value} $\wp( \pi )$ of $\pi$ is the profile which counts
 how often a color $k\in \{0,1,\ldots, d-1\}$ appears in $c(s_0) c(s_1) \ldots c(s_l)$.
For an infinite sequence $\pi = s_0 s_1 \ldots$, 
 its {\em value} $\wp( \pi )$ is defined to be $\infty$, if $\pi$
 is won by player $0$ w.r.t. the parity game winning condition; otherwise $\wp(\pi) := -\infty$.
Finally, we introduce a total order $\prec$ on $\mathcal{P}$ which tries to capture the notion
of when one of two given plays is better than the other for player $0$:
For this we set (i) $-\infty$ to be the bottom element of $\prec$, 
(ii) $\infty$ to be the top element of $\prec$, and (iii)
for all $\wp,\wp' \in \mathcal{P}\setminus\{-\infty,\infty\}$ we set:
\[
\begin{array}{lcrl}
	\wp \prec \wp' & :\Leftrightarrow & \exists k\in \{0,1,\ldots,d-1\} : & k = \max\{ k\in\{0,1,\ldots,d-1\} \mid \wp_k \neq \wp'_k \}\\
	               & & \wedge & \left( k \equiv_2 0 \wedge \wp_k < \wp'_k \vee k \equiv_2 1 \wedge \wp_k > \wp'_k \right).
\end{array}
\]
Informally, the definition of $\prec$ says that player $0$ hates to loose in an infinite play,
whereas he likes it the most to win an infinite play. So, whenever he can, he will try to escape
an infinite play he cannot win, therefore resulting in a finite play to $\bot$: here, given two finite plays $\pi_1,\pi_2$ ending in $\bot$, player $0$ looks for the highest color $c$ which does not appear 
equally often along both plays. If $c$ is even, he prefers that play in which it appears more often;
if it is odd, he prefers the one in which it appears less often. 
In particular, player $0$ dislikes visiting odd-dominated cycle, while he likes visiting even-dominated ones:
\begin{lemma}\label{lem:cycle-weight}
Assume that $\chi = s_0 s_1 \ldots s_n$ is a non-empty cycle in the parity game arena $\mathcal{A}$, i.e. $s_0 \in s_n E$ and $n \ge 0$. 
$\chi$ is $0$-dominated, i.e. the highest color in $\chi$ is even if and only if $\wp(\chi) \succ \zp$.
$\chi$ is $1$-dominated if and only if $\wp(\chi) \prec \zp$.
\end{lemma}
Now, for a given parity game arena $\A$ let $\sigma^\ast_0$, 
$\sigma^\ast_1$ be the optimal winning strategies of player $0$, resp. $1$. Further, let $W_0$, $W_1$ be the corresponding winning sets.
Obviously, both players can still use these strategies in $\Ab$, 
too, as we only added additional edges. 
Especially, player $0$ can still use $\sigma^\ast_0$ to win $W_0$ in $\Ab$ as only he has the option
to move to $\bot$. 
In the case of player $1$, by applying $\sigma^\ast_1$ any cycle in $\Ab|_{\sigma^\ast_1}$ reachable
from a vertex $v\in W_1$ has to be odd-dominated. 
Hence, player $0$ prefers to play in an acyclic path from $v$ to $\bot$ in $\Ab|_{\sigma_1}$ when starting from a vertex in $W_1$.

Let therefore be $\overline{\wp}$ the $\prec$-maximal value of any acyclic path terminating in $\bot$ in $\Ab$. 
$\overline{\wp}$ is the best player $0$ can hope to achieve starting from a node $v\in W_1$ when player $1$ plays
optimal.
We therefore define: player $0$ wins a play $\pi$, if $\wp(\pi)\succ \overline{\wp}$, otherwise player $1$ wins the play. Player $i$ wins a node $s\in V$, if he has a strategy $\sigma \subseteq E_i$ 
with which he wins any play starting from $s$ in $\A|_{\sigma}$.
As already sketched, this leads then to the following theorem.
\begin{theorem}\label{thm:w0-unchanged}
Player $i$ wins the node $s$ in $\mathcal{A}$ iff he wins it in $\mathcal{A}_\bot$.
\end{theorem}

\section{Strategy Improvement}
We now turn to the problem of finding optimal winning strategies by iteratively
valuating the strategy, and determining from this valuation possible better strategies.
The following section can be seen as the generalization of the algorithm
in \cite{BSV04} to non-deterministic strategies 
and explicitly stated in the setting of parity games.
In fact, we will only consider a special class of strategies for player $0$,
i.e. such strategies which do not introduce any $1$-dominated cycles.
The strategy improvement process will assure that no $1$-dominated cycles are created.
If there are any $1$-dominated cycles in $\mathcal{A}_\bot |_{V_1}$,
then player $1$ wins all the nodes in the $1$-attractor to these cycles.
We may, thus, identify the nodes trivially won by player $1$ in a preprocessing step, and remove them.
\begin{ass}
 The arena $\mathcal{A}_\bot |_{V_1}$ has no $1$-dominated cycles.
\end{ass}
\begin{definition}
We call a strategy $\sigma \subseteq E_0$ of player $0$ {\em reasonable},
if there are no $1$-dominated cycles in $\mathcal{A}_\bot |_{\sigma}$.
\end{definition}
\begin{remark}
{\bf (a)}
By our assumption above the strategy $\sigma_\bot := V_0 \times \{\bot\}$ is reasonable, 
as every $1$-dominated cycle in $\A$
consists of at least one node controlled by player $0$.
{\bf (b)}
Let $\sigma$ be any strategy of player $0$, and $W_\sigma$ the set of nodes won by $\sigma$.
Then, the strategy 
$\sigma' = \sigma \cap ( W_\sigma \times W_{\sigma}) \cup \{ (s,\bot) \mid s \in V_0 \setminus W_\sigma \}$
is reasonable with $W_{\sigma} = W_{\sigma'}$.
\end{remark}
We may thus assume that player $0$ uses only reasonable strategies.
\begin{definition}\label{def:val}
Let $\sigma$ be some reasonable strategy of player $0$.
Its valuation $\vals: V\cup\{\bot\} \to \mathcal{P}$ maps
every node $s$ on the $\prec$-minimal value $\vals(s)$
which player $1$ can {\em guarantee} to achieve in any play starting from $s$ in $\Ab|_{\sigma}$ by using some 
memoryless strategy:
\[
	\vals(s) 
	:= \min^\prec_{\tau \subseteq E_1 \text{ strategy}} \max^\prec\{ \wp(\pi) \mid \pi \text{ is a play in } \Ab|_{\sigma,\tau} \wedge \pi(0) = s \},
\]
where we set $\vals(\bot) :=\zp$.
\end{definition}
\begin{remark}
{\bf (a)} We will show later that, if we start from the reasonable strategy $\sigma_{\bot} := V_0 \times\{\bot\}$ , then
our strategy-improvement algorithm will only generate reasonable strategies.
(Note, if $\Ab|_{\sigma_\bot}$ had $1$-dominated cycles, 
then these would need to exist solely in $\A|_{V_1}$ 
-- but we have assumed above that we removed those in a preprocessing step.)
{\bf (b)} As shown above, for all $s\in W_1$ player $1$ can use his optimal winning strategy $\sigma^\ast_1$ from the parity game
to guarantee $\vals(s) \preceq \overline{\wp} \prec \infty$.
\end{remark}
By means of the valuation $\vals$ we can partially order reasonable strategies in the natural way:
\begin{definition}
For two (reasonable) strategies $\sigma_a, \sigma_b$ of player $0$ we write
$\sigma_a \preceq \sigma_b$, if $\val_{\sigma_a}(s) \preceq \val_{\sigma_b}(s)$ for all nodes $s$.
We write $\sigma_a \prec \sigma_b$, if there is at least one node $s$ such that $\val_{\sigma_a}(s) \prec \val_{\sigma_b}(s)$.
Finally, $\sigma_a \approx \sigma_b$, if $\sigma_a\preceq \sigma_b \wedge \sigma_b\preceq \sigma_a$.
\end{definition}

\noindent The following lemma addresses the calculation of $\vals$ using a straight-forward adaption of the Bellman-Ford algorithm:
\begin{lemma}\label{lem:bellman-ford}
Let $\sigma\subseteq E_0$ be a reasonable strategy of player $0$.
We define $\val_{\bot}: V\cup\{\bot\} \to \mathcal{P}$ 
by $\val_{\bot}(\bot) := \zp$, and $\val_{\bot}(s) = \infty$ for all $s\in V$,
and the operator $F_\sigma : (V\cup\{\bot\} \to \mathcal{P}) \to (V\cup\{\bot\} \to \mathcal{P})$ by
\[
    \begin{array}{lcll}
	 	F_{\sigma}[\val](\bot) & := & \zp &\\
	 	F_{\sigma}[\val](s)    & := & \wp(s) + \min^{\preceq} \{ \val(t) \mid (s,t)\in E_1 \} & \text{ if } s\in V_1,\\
	 	F_{\sigma}[\val](s)    & := & \wp(s) + \max^{\preceq} \{ \val(t) \mid (s,t)\in \sigma \} & \text{ if } s\in V_0,
		\end{array}
\]
for any $\val : V\cup\{\bot\} \to \mathcal{P}$.

Then, the valuation $\vals$ of $\sigma$ is given
as the limit of the sequence $F^i_\sigma[\val_\bot]$ for $i\to \infty$, 
and this limit is reached after at most $\abs{V}$ iterations.
\end{lemma}
\begin{remark}
{\bf (a)} We assume unit cost for adding and comparing color profiles.
The time needed for calculating $\vals$ is then simply given
by $O(\abs{V}\cdot \abs{E})$.%
{\bf (b)} For every $s\in V$ 
    there has to be at least one edge $(s,t)$ with $\vals(s) = \wp(s) + \vals(t)$,
    as $\vals = F_{\sigma}[\vals]$.
\end{remark}
W.r.t. $\vals$ we can identify possible {\em improvements} of $\sigma$: 
\begin{definition}
Let $\sigma \subseteq E_0$ be a reasonable strategy of player $0$.
The set $\sopt$ of {\em improvements}, resp. the set $S_{\sigma}$ of {\em strict improvements} of $\sigma$ is
defined by
\begin{itemize}
\item $\sopt := \{ (s,t)\in E_0 \mid \vals(s) \preceq \wp(s) + \vals(t) \}$, resp.\
\item $S_{\sigma} := \{ (s,t)\in E_0 \mid \vals(s) \prec \wp(s) + \vals(t) \}$.
\end{itemize}
We call a strategy $\sigma' \subseteq E_0$ a {\em direct improvement} of
$\sigma$ if $\sigma'\subseteq \sopt$.
\end{definition}

\begin{fact}
Let $\sigma'$ be a direct improvement of $\sigma$. 
Then along every edge $(u,v)$ of $\mathcal{A}_\bot |_{\sigma'}$
we have 
$\val_{\sigma}(u) \preceq \wp(u) + \val_{\sigma}(v)$.
In particular, we have for any finite path $s_0 s_1 \ldots s_{l+1}$ in $\Ab|_{\sigma'}$
\[
	\vals( s_0 ) \preceq \wp(s_0) + \vals(s_1) \preceq \wp( s_0 s_1 ) + \vals(s_2) \preceq \ldots \preceq \wp(s_0\ldots s_l ) + \vals(s_{l+1}).
\]
\end{fact}
From this easy fact, several important properties of direct improvements follow:
\begin{corollary}\label{cor:reasonable-impr}
If $\sigma$ is reasonable, then any $0$-strategy $\sigma'\subseteq I_{\sigma}$ is reasonable, too.
\end{corollary}

\begin{corollary}\label{cor:impr-improves}
Let $\sigma$ be a reasonable strategy.
For a direct improvement $\sigma'$ of $\sigma$ we have that 
$\sigma \preceq \sigma'$.
If $\sigma'$ contains at least one strict improvement of $\sigma$, 
then this inequality is strict, i.e. ${\sigma} \prec {\sigma'}$.
\end{corollary}
The preceding corollaries show that starting with an initial reasonable strategy $\sigma_0$, e.g. $\sigma_{\bot}$,
we can generate a sequence $\sigma_0, \sigma_1, \sigma_2, \ldots$
of reasonable strategies such that $\val_{\sigma_i}(s) \preceq \val_{\sigma_{i+1}}(s)$ for all $s\in V$,
if we choose the strategy $\sigma_{i+1}$ to  be some direct improvement of $\sigma_i$.
Further, we know, if  $\sigma_{i+1}$ uses at least one strict improvement $(s,t)$ of $\sigma_i$, i.e. $(s,t)\in\sigma_{i+1}\cap S_{\sigma_i} \neq \emptyset$, then we have $\val_{\sigma_i}(s) \prec \val_{\sigma_{i+1}}(s)$,
i.e. every possible reasonable strategy occurs at most once along the strategy improvement sequence.
As already shown, we have always $\val_{\sigma_i}(s) \preceq \overline{\wp} \prec \infty$ for all nodes $s\in W_1$.
The obvious question is now, if we can reach an optimal winning strategy by this procedure, i.e.
is a reasonable strategy $\sigma$ with $S_{\sigma} = \emptyset$ optimal?
This is answered in the following lemma.
\begin{lemma}\label{lem:impr-till-optimal}
As long as there is a node $s\in W_0$ with $\val_{\sigma}(s) \prec \infty$, $\sigma$ has at least one strict improvement.
\end{lemma}
Due to this lemma, we know that, if a reasonable strategy $\sigma$ has no strict improvements,
i.e. $S_{\sigma}=\emptyset$, then we have $\vals(s) = \infty$ for at least all the nodes $s\in W_0$.
On the other hand, for all nodes $s\in W_1$ we always have $\vals(s) \preceq \overline{\wp}$.
Hence, by the determinacy of parity games, i.e. $W_1 = V \setminus W_0$, $\sigma$ has to be an optimal
winning strategy for player $0$, if $S_{\sigma} = \emptyset$.
By our construction such an optimal strategy $\sigma$ with $S_\sigma = \emptyset$ might be non-deterministic.
The following lemma shows how one can deduce an optimal deterministic strategy from such a $\sigma$.
\begin{lemma}\label{lem:deterministic-optimal}
Let $\sigma$ be a reasonable strategy of player $0$ in $\Ab$, \
and $\sopt$
the strategy consisting of all improvements of $\sigma$.
Then every deterministic strategy $\sigma'\subseteq \sopt$ with $\val_{\sopt}(s) = \wp(s) + \val_{\sopt}(t)$
for all $(s,t) \in \sigma'$ satisfies $\val_{\sopt} = \val_{\sigma'}$.
\end{lemma}
Starting from $\sigma_\bot = \{ (s,\bot) | s\in V_0\}$, 
if we improve the current strategy using at least one strict improvement in every step, we
will end up with an optimal winning strategy for player $0$. 
As in every step the valuation increases
in at least those nodes at which a strict improvement exists, 
and as there are at most $( \frac{\abs{V}}{d} + 1)^d$ possible values a valuation
can assign a given node,
the number of improvement steps is bound
by $\abs{V} \cdot ( \frac{\abs{V}}{d} + 1 )^{d}$.
The cost of every improvement step is given by the cost of the calculation of $\val_{\sigma}$, we thus get:
\begin{theorem}
Let $\sigma_0$ be some reasonable $0$-strategy. 
By iteratively taking $\sigma_{i+1}$ to be some direct improvement of $\sigma_i$ which uses
at least one strict improvement, one obtains an optimal winning strategy after at most 
$\abs{V} \cdot ( \frac{\abs{V}}{d} + 1 )^{d}$ iterations. The total running time is thus
 $O( \abs{V}^2 \cdot \abs{E} \cdot ( \frac{\abs{V}}{d} + 1 )^{d} )$.
\end{theorem}

\subsection{All Profitable Switches}
In the previous subsection we have not said anything about which direct improvement 
should be taken in every improvement step.
As no algorithms are known which determine for a given strategy
such a direct improvement that the total number of improvement steps is minimal (we call
such a direct improvement ``globally optimal''),
one usual resorts to heuristics for choosing a direct improvement,
(see e.g. \cite{BSV02}).% for a comparison of different heuristics).
Most often the heuristic ``all profitable switches'' mentioned in the introduction is used.
In the case of non-deterministic strategies this simply becomes taking $I_\sigma$ as successor 
strategy. The interesting fact here is that $I_\sigma$ is  
a ``locally optimal'' direct improvement for 
a given reasonable strategy $\sigma$, i.e. for all strategies $\sigma' \subseteq \sopt$
we have $\sigma' \preceq I_\sigma$.
We remark that this has already been shown implicitly by Schewe in \cite{S07b}:
\begin{theorem}
Let $\sigma$ be a reasonable strategy with $I_{\sigma}$ its set of improvements.
For any direct improvement of $\sigma$ we have $\sigma' \preceq I_{\sigma}$.%,
\end{theorem}
We like to give an easy proof for this theorem. We first note the following 
two properties of the operator $F_{\sigma}$:
\begin{fact}
{\bf (i)} For $\val,\val' : V\cup\{\bot\} \to \mathcal{P}$  
with $\val \preceq \val'$ we have $F_{\sigma}[\val] \preceq F_{\sigma}[\val']$.\\
{\bf (ii)} For two $0$-strategies $\sigma_a \subseteq \sigma_b$ 
we have $F_{\sigma_a}[\val](s) \preceq F_{\sigma_b}[\val](s)$ for all $s\in V$.
\end{fact}
Using (i) and (ii) we get by induction
\[
F^{i+1}_{\sigma_{a}}[\val_\bot] = F_{\sigma_a}[ F^i_{\sigma_a}[\val_\bot] ] \preceq F_{\sigma_a}[ F^i_{\sigma_b}[\val_\bot]] \preceq F_{\sigma_b}[ F^i_{\sigma_b}[\val_\bot]] = F^{i+1}_{\sigma_b}[ \val_\bot ],
\]
and therefore the following lemma:
\begin{lemma}
If $\sigma_a$ and $\sigma_b$ are reasonable and $\sigma_a\subseteq \sigma_b$, it holds that $\val_{\sigma_a} \preceq \val_{\sigma_b}$.
\end{lemma}
Now, as the set of improvements $I_\sigma$ of a given reasonable strategy $\sigma$ is itself
a (non-deterministic) strategy, and every direct improvement $\sigma'$ of $\sigma$ satisfies $\sigma'\subseteq I_{\sigma}$
by definition, the theorem from above follows. The algorithm of Schewe in \cite{S07b} can therefore
be described as an optimized implementation of non-deterministic strategy iteration using the ``all profitable switches'' heuristic.

We close this section with a remark on the calculation of $\val_{I_\sigma}$.
Schewe proposes an algorithm for calculating $\val_{I_\sigma}$ 
which uses $\vals$ to speed up the calculation leading to
 $O(\abs{E}\log \abs{V})$ operations on color-profiles instead of $O(\abs{E}\cdot\abs{V})$.
For this, formulated in the notation of our algorithm,
he introduces edge weights $w(u,v) := (\wp(u)+ \vals(v)) - \vals(u)$,
and calculates w.r.t. these edges an update $\delta = \val_{I_\sigma} - \vals$.
We argue that one can use Dijkstra's algorithm for this,
as we have $\vals(u) \preceq \wp(u) + \vals(v)$ along all
edges $(u,v)\in I_{\sigma}$, and thus $w(u,v)\succeq \zp$,
i.e. all edge weights are non-negative.
\begin{proposition}\label{prop:dij}
$\val_{I_{\sigma}}$ can be calculated using Dijkstra's algorithm which needs $O(\abs{V}^2)$
operations on color-profiles on dense graphs;
for graphs whose out-degree is bound by some $b$ this can be improved to $O(b\cdot \abs{V} \cdot \log \abs{V})$ by
using a heap
\footnote{
In \cite{BSV04} the authors propose another optimization to speed up the calculation of $\vals$ 
by restricting the re-calculation of $\vals$ to only those nodes where $\vals$ changes.
Those nodes can be easily identified by calculating an attractor again in time $O(\abs{E})$.
Unfortunately, combining this optimization with the one by Schewe (\cite{S07b}) does not lead to a better
asymptotic upper bound.
}. 
\end{proposition}

\noindent This gives us a running time of $O( \abs{V}^3 \cdot ( \frac{\abs{V}}{d} + 1 )^{d} )$, resp. $O( \abs{V}^2 \cdot b\cdot \log\abs{V}\cdot ( \frac{\abs{V}}{d} + 1 )^{d} )$.% as an upper bound.

\section{Comparison with the Algorithm by Jurdzinski and V{\"o}ge}
This section compares the algorithm presented in this article 
with the one by Jurdzinski and V{\"o}ge \cite{JV00}.
We first give a short (slightly imprecise) description of the algorithm in \cite{JV00}:
This algorithm starts in each step with some {\em deterministic} $0$-strategy $\sigma$.
Using $\sigma$ a valuation $\Omega_\sigma$ is calculated 
(see below for details about $\Omega_\sigma$).
Then, by means of this valuation possible strategy improvements are determined, and finally some 
non-empty subset of these improvements is chosen, but only one improvement per node at most,
such that implementing these improvements yields a {\em deterministic} strategy again.
This process is repeated until there are no improvements anymore w.r.t. the current strategy.

\paragraph{The valuation $\Omega_{\sigma}$:} We present a slightly ``optimized'' version
of the valuation used in \cite{JV00}.
The valuation $\Omega_\sigma(s)$ of a deterministic $0$-strategy $\sigma$
consists of the the {\em cycle value} $z_\sigma(s)$, the {\em path profile} $\wp_\sigma(s)$, 
and the {\em path length} $l_\sigma(s)$ which are defined as follows:
\begin{itemize}
\item
  As $\sigma$ is deterministic, all plays in $\A|_{\sigma}$ are determined by player $1$.
  For every node $z$ having odd color, we can decide whether there is at least 
  one cycle in $\A|_{\sigma}$ such that this cycle is dominated by $z$.
  Let $Z$ be the set of all odd colored nodes dominating a cycle in $\A|_{\sigma}$.
  
  Given a node $s$ we define $z_\sigma(s)$ to be a node of maximal color in $Z$,
  which is reachable from $s$ in $\A|_{\sigma}$; if no node in $Z$ is reachable
  from $s$ in $\A|_{\sigma}$, then $s$ has to be won by player $0$, and we 
  set $z_{\sigma}(s) = \infty$.
\item
  If $z_{\sigma}(s)$ is some odd colored node, 
	the second component $\wp_{\sigma}(s)$ becomes the color profile of a $\prec$-minimal play from $s$
  to $z_\sigma(s)$ in $\A|_{\sigma}$
  -- with the restriction that only nodes of color $\ge c(z_\sigma(s))$ are counted.
\item
  Finally, if $\wp_{\sigma}(s)$ is defined, $l_{\sigma}(s)$ is the length of shortest play from $s$ to $z_\sigma(s)$ w.r.t. $\wp_\sigma(s)$, if $z_\sigma(s)$ has odd color.
\end{itemize}
\begin{remark}
We assume here that $z_{\sigma}$ is either $\infty$, if $s$ is already won using $\sigma$, or the ``worst'' odd-dominated
cycle into which player $1$ can force a play starting from $s$. 
In \cite{JV00}, the authors even try to optimize $z_{\sigma}(s)$ when $s$ is already won using $\sigma$.
These improvements are obviously unnecessary, as we can always remove the attractor to these nodes from
the arena in an intermediate step in order to obtain a smaller arena.

Further, it is assumed in \cite{JV00} that every node is uniquely colored.
Therefore,  in \cite{JV00} $\wp_{\sigma}$ is defined to be the {\em set} of nodes having higher color 
than $z_{\sigma}(s)$ on a ``worst'' path from $s$ to $z_{\sigma}(s)$. 
Jurdzinski and V{\"o}ge already mention at the end of \cite{JV00} that their algorithm
also works when not assuming that every vertex is uniquely colored, but do not
present the adapted data structures needed in this case. This was done in \cite{BSV03}: 
If the same color is used for several vertices, 
it is sufficient to only count the number of nodes having a color $k\ge z_{\sigma}(s)$
along such a ``worst'' path from $s$ to $z_{\sigma}(s)$ where ``worst'' path simply means a $\prec$-minimal path then. Therefore,
the color profiles used in this article are a direct generalization of the path profiles
used in \cite{JV00}.
\end{remark}
In \cite{JV00} an edge $(s,t)\in E_0$ is now called a strict improvement over $(s,\sigma(s))$,
if $\Omega_\sigma(t)$ is strictly better than $\Omega_{\sigma}(\sigma(s))$, i.e.
either the ``worst'' cycle improves, or the worst play to it improves, or the length of 
a worst play becomes longer (``the longer player $0$ can stay away from $z_{\sigma}$ the better for him'').
A deterministic strategy $\sigma'$ is then a direct improvement of a given deterministic strategy $\sigma$
w.r.t. \cite{JV00}, if it differs from $\sigma$ only in strict improvements.

\begin{definition}
For a given parity game arena $\A = (V,E,c,o)$, set
\[
	\A^{\bot} := ( V \cup\{\bot\}, E\cup V_0 \times \{\bot\} \cup \{(\bot,\bot)\}, c \cup \{ (\bot, -1) \}, o\cup\{ (\bot,0)\} ).
\]
\end{definition}
$\A^{\bot}$ results from $\Ab$ by simply adding a loop to $\bot$, and giving $\bot$  the color $-1$
so that $\A^{\bot}$ is a parity game arena where $\bot$ is the cycle dominated by
the least odd color.
A straight-forward adaption of the proof of Theorem~\ref{thm:w0-unchanged} shows that player $0$ wins a node $s$ in $\A$ iff he wins it in $\A^{\bot}$.

Now, as the strategy improvement algorithm in \cite{JV00} tries to play to the ``best''
possible cycle, an optimal strategy (obtained by the algorithm) will always choose to play to $\bot$ from a node $s$,
if $s$ cannot be won by player $0$, as every other $1$-dominated cycle has at least $1$ as maximal color.
A strategy $\sigma$ of player $0$ is therefore ``reasonable'' w.r.t. to the algorithm
by Jurdzinski and V{\"o}ge, if $(\bot,\bot)$ is the only $1$-dominated cycle in $\A^{\bot}|_{\sigma}$.

Obviously, we now have an one-to-one correspondence between reasonable strategies $\sigma$ in $\Ab$,
and reasonable strategies $\sigma$ in $\A^{\bot}$ of player $0$: we simply have to remove or add the edge $(\bot,\bot)$
to move from $\Ab$ to $\A^{\bot}$ and vice versa. 
We therefore may identify these strategies in the following
as one strategy.

This allows us to compare the improvement step
of the algorithm presented in this article with that of \cite{JV00}.
Indeed, as the color of $\bot$ is $-1$ (recall that all other nodes
have colors $\ge 0$), we have $\wp_{\sigma}(s) = \vals(s)$ for 
all nodes with $z_\sigma(s) = \bot$, and $\vals(s) = \infty$, if $z_\sigma(s) \neq \bot$.
This proves the following proposition:
\begin{proposition}
Any (deterministic) direct improvement $\sigma'$ of $\sigma$ identified by \cite{JV00} is  
a subset of $I_{\sigma}$.
Therefore $\sigma' \preceq I_{\sigma}$.
\end{proposition}
In other words, the algorithm presented here always chooses
{\em locally} a direct improvement of $\sigma$ which is at least as good
as any deterministic direct improvement obtainable by \cite{JV00}. 
In the appendix, a small example can be found illustrating this.

\subsection{Bound on the number of Improvement Steps}
We finish this section by giving an upper bound on the total number of improvement steps 
when using the ``all profitable switches''-heuristic.
In the case of an arena with out-degree two, % 
one can show that the
number of improvement steps done by 
the algorithm in \cite{JV00} is bounded by $O(\frac{2^{\abs{V_0}}}{\abs{V_0}})$ (cf. \cite{BSV02}).

When considering non-deterministic strategies the heuristic ``all profitable switches''
naturally generalizes to simply taking $I_{\sigma}$ as successor strategy in every iteration.
Here we can show the following upper bound:
\begin{theorem}\label{thm:steps}
Let $\Ab$ be a escape-parity-game arena where every node of player $0$ has at most
two successor. Then the number of improvement steps needed to reach an optimal winning
strategy is bound by $3\cdot 1.724^{\abs{V_0}}$ when using non-deterministic strategy iteration
and the ``all profitable switches''-heuristic.
\end{theorem}
\begin{remark}
To the best of our knowledge this is the best upper bound known for any
deterministic strategy-improvement algorithm. In \cite{BSV02} a similar
bound is only obtained by using randomization.
\end{remark}

\section{Conclusions}
In the first part of the article,
we presented an extended version of the algorithm by \cite{BSV04}
which (i) allows the use of non-deterministic strategies,
and (ii) works directly on the given parity game arena without requiring
a reduction to a mean payoff game as an intermediate step.
For (ii), we used the path profiles introduced in \cite{JV00},
resp. a generalized version of it called color profiles (see also \cite{BSV03}).

We then showed that the heuristic ``all profitable switches'' in the setting of 
non-deterministic strategies leads to the locally best direct improvement,
and therefore to the algorithm presented in \cite{S07b}.%
We further identified the fast calculation of the valuation proposed by Schewe 
as the Dijkstra algorithm.

Finally, we turned to the comparison of the algorithm presented here 
to the one by Jurdzinski and V{\"o}ge \cite{JV00}.
As our algorithm works directly on parity games in contrast to \cite{BSV04,S07b}, 
we could show that the valuations used in both coincide for parity game arenas
with escape for player $0$. 
We finished the article by adapting results from \cite{MS99} which allowed us to 
show that using the ``all profitable switches''-heuristic in the setting of non-deterministic strategies allows to obtain an upper bound of $O(1.724^{\abs{V_0}})$
on the total number of improvement steps. This bound also carries over to the algorithm in \cite{S07b}. 
This bound was previously only attainable using randomization \cite{BSV02}.
\bibliographystyle{plain} %oder alpha oder splncs
\bibliography{db}

\begin{thebibliography}{10}

\bibitem{BSV02}
H.~Bj{\"o}rklund, S.~Sandberg, and S.~Vorobyov.
\newblock Optimization on completely unimodal hypercubes.
\newblock Technical Report 2002--18, Department of Information Technology,
  Uppsala University, 2002.

\bibitem{BSV03}
H.~Bj{\"o}rklund, S.~Sandberg, and S.~Vorobyov.
\newblock A discrete subexponential algorithm for parity games.
\newblock In {\em STACS'03}, LNCS 2607, pages 663--674. Springer, 2003.

\bibitem{BSV04}
H.~Bj{\"o}rklund, S.~Sandberg, and S.~Vorobyov.
\newblock A combinatorial strongly subexponential strategy improvement
  algorithm for mean payoff games.
\newblock In {\em MFCS'04}, LNCS 3153, pages 673--685. Springer, 2004.

\bibitem{EJ91}
E.A. Emerson and C.S. Jutla.
\newblock Tree automata, mu-calculus and determinacy (extended abstract).
\newblock In {\em FOCS'91}. IEEE Computer Society Press, 1991.

\bibitem{HK66}
A.~Hoffman and R.~Karp.
\newblock On nonterminating stochastic games.
\newblock {\em Management Science}, 12, 1966.

\bibitem{How60}
Ronald~A. Howard.
\newblock Dynamic programming and markov processes.
\newblock {\em The M.I.T. Press}, 1960.

\bibitem{JPZ06}
M.~Jurdzi{\'n}ski, M.~Paterson, and U.~Zwick.
\newblock A deterministic subexponential algorithm for solving parity games.
\newblock In {\em SODA'06}. ACM/SIAM, 2006.

\bibitem{Jur98:IPL}
Marcin Jurdzi{\'n}ski.
\newblock Deciding the winner in parity games is in {UP}~$\cap$~co-{UP}.
\newblock {\em Information Processing Letters}, 68(3):119--124, November 1998.

\bibitem{Jur00:STACS}
Marcin Jurdzi{\'n}ski.
\newblock Small progress measures for solving parity games.
\newblock In {\em STACS 2000}, volume 1770 of {\em LNCS}, 2000.

\bibitem{MS99}
Y.~Mansour and S.~Singh.
\newblock On the complexity of policy iteration.
\newblock In {\em UAI 1999}, 1999.

\bibitem{Mos91}
A.W. Mostowski.
\newblock Games with forbidden positions.
\newblock Technical Report~78, University of Gda\'{n}sk, 1991.

\bibitem{Puri}
Anuj Puri.
\newblock {\em Theory of Hybrid Systems and Discrete Event Systems}.
\newblock PhD thesis, Electronic Research Laboratory, College of Engineering,
  University of California, Berkley, 1995.

\bibitem{S07b}
Sven Schewe.
\newblock An optimal strategy improvement algorithm for solving parity games.
\newblock Technical Report~28, Universit{\"a}t Saarbr{\"u}cken, 2007.

\bibitem{GS07:CSL}
H.~Seidl and T.~Gawlitza.
\newblock Precise relational invariants through strategy iteration.
\newblock In {\em CSL'07}, LNCS, 2007.

\bibitem{JV00}
Jens V{\"{o}}ge and Marcin Jurdzi{\'{n}}ski.
\newblock A discrete strategy improvement algorithm for solving parity games
  ({E}xtended abstract).
\newblock In {\em CAV'00}, volume 1855 of {\em LNCS}, 2000.

\end{thebibliography}
\newpage
\appendix
\section{Example: Comparison with the Algorithm by Jurdzinski and V{\"o}ge}

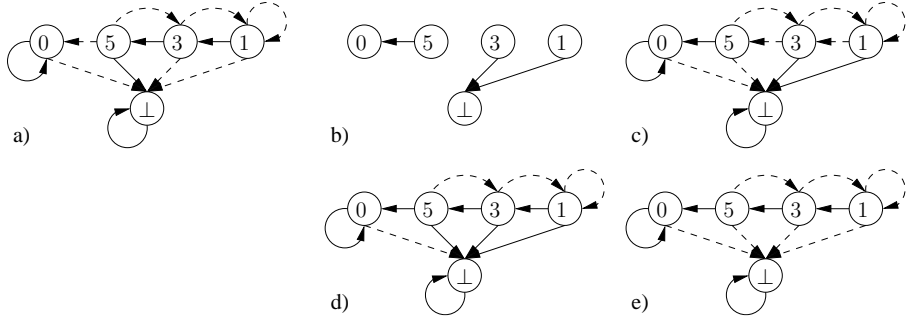
\begin{center}\scalebox{0.7}{\begin{picture}(0,0)%
\includegraphics{ex-arena-2.pstex}%
\end{picture}%
\setlength{\unitlength}{3947sp}%
\begingroup\makeatletter\ifx\SetFigFont\undefined%
\gdef\SetFigFont#1#2#3#4#5{%
  \reset@font\fontsize{#1}{#2pt}%
  \fontfamily{#3}\fontseries{#4}\fontshape{#5}%
  \selectfont}%
\fi\endgroup%
\begin{picture}(8071,2821)(90,-2071)
\put(3001,-2011){\makebox(0,0)[lb]{\smash{{\SetFigFont{12}{14.4}{\familydefault}{\mddefault}{\updefault}{\color[rgb]{0,0,0}d)}%
}}}}
\put(5926,314){\makebox(0,0)[lb]{\smash{{\SetFigFont{12}{14.4}{\familydefault}{\mddefault}{\updefault}{\color[rgb]{0,0,0}$0$}%
}}}}
\put(6526,314){\makebox(0,0)[lb]{\smash{{\SetFigFont{12}{14.4}{\familydefault}{\mddefault}{\updefault}{\color[rgb]{0,0,0}$5$}%
}}}}
\put(7126,314){\makebox(0,0)[lb]{\smash{{\SetFigFont{12}{14.4}{\familydefault}{\mddefault}{\updefault}{\color[rgb]{0,0,0}$3$}%
}}}}
\put(7726,314){\makebox(0,0)[lb]{\smash{{\SetFigFont{12}{14.4}{\familydefault}{\mddefault}{\updefault}{\color[rgb]{0,0,0}$1$}%
}}}}
\put(6826,-286){\makebox(0,0)[lb]{\smash{{\SetFigFont{12}{14.4}{\familydefault}{\mddefault}{\updefault}{\color[rgb]{0,0,0}$\bot$}%
}}}}
\put(5701,-511){\makebox(0,0)[lb]{\smash{{\SetFigFont{12}{14.4}{\familydefault}{\mddefault}{\updefault}{\color[rgb]{0,0,0}c)}%
}}}}
\put(376,314){\makebox(0,0)[lb]{\smash{{\SetFigFont{12}{14.4}{\familydefault}{\mddefault}{\updefault}{\color[rgb]{0,0,0}$0$}%
}}}}
\put(976,314){\makebox(0,0)[lb]{\smash{{\SetFigFont{12}{14.4}{\familydefault}{\mddefault}{\updefault}{\color[rgb]{0,0,0}$5$}%
}}}}
\put(1576,314){\makebox(0,0)[lb]{\smash{{\SetFigFont{12}{14.4}{\familydefault}{\mddefault}{\updefault}{\color[rgb]{0,0,0}$3$}%
}}}}
\put(2176,314){\makebox(0,0)[lb]{\smash{{\SetFigFont{12}{14.4}{\familydefault}{\mddefault}{\updefault}{\color[rgb]{0,0,0}$1$}%
}}}}
\put(1276,-286){\makebox(0,0)[lb]{\smash{{\SetFigFont{12}{14.4}{\familydefault}{\mddefault}{\updefault}{\color[rgb]{0,0,0}$\bot$}%
}}}}
\put(151,-511){\makebox(0,0)[lb]{\smash{{\SetFigFont{12}{14.4}{\familydefault}{\mddefault}{\updefault}{\color[rgb]{0,0,0}a)}%
}}}}
\put(3226,314){\makebox(0,0)[lb]{\smash{{\SetFigFont{12}{14.4}{\familydefault}{\mddefault}{\updefault}{\color[rgb]{0,0,0}$0$}%
}}}}
\put(3826,314){\makebox(0,0)[lb]{\smash{{\SetFigFont{12}{14.4}{\familydefault}{\mddefault}{\updefault}{\color[rgb]{0,0,0}$5$}%
}}}}
\put(4426,314){\makebox(0,0)[lb]{\smash{{\SetFigFont{12}{14.4}{\familydefault}{\mddefault}{\updefault}{\color[rgb]{0,0,0}$3$}%
}}}}
\put(4126,-286){\makebox(0,0)[lb]{\smash{{\SetFigFont{12}{14.4}{\familydefault}{\mddefault}{\updefault}{\color[rgb]{0,0,0}$\bot$}%
}}}}
\put(5026,314){\makebox(0,0)[lb]{\smash{{\SetFigFont{12}{14.4}{\familydefault}{\mddefault}{\updefault}{\color[rgb]{0,0,0}$1$}%
}}}}
\put(3001,-511){\makebox(0,0)[lb]{\smash{{\SetFigFont{12}{14.4}{\familydefault}{\mddefault}{\updefault}{\color[rgb]{0,0,0}b)}%
}}}}
\put(5926,-1186){\makebox(0,0)[lb]{\smash{{\SetFigFont{12}{14.4}{\familydefault}{\mddefault}{\updefault}{\color[rgb]{0,0,0}$0$}%
}}}}
\put(6526,-1186){\makebox(0,0)[lb]{\smash{{\SetFigFont{12}{14.4}{\familydefault}{\mddefault}{\updefault}{\color[rgb]{0,0,0}$5$}%
}}}}
\put(7126,-1186){\makebox(0,0)[lb]{\smash{{\SetFigFont{12}{14.4}{\familydefault}{\mddefault}{\updefault}{\color[rgb]{0,0,0}$3$}%
}}}}
\put(7726,-1186){\makebox(0,0)[lb]{\smash{{\SetFigFont{12}{14.4}{\familydefault}{\mddefault}{\updefault}{\color[rgb]{0,0,0}$1$}%
}}}}
\put(6826,-1786){\makebox(0,0)[lb]{\smash{{\SetFigFont{12}{14.4}{\familydefault}{\mddefault}{\updefault}{\color[rgb]{0,0,0}$\bot$}%
}}}}
\put(5701,-2011){\makebox(0,0)[lb]{\smash{{\SetFigFont{12}{14.4}{\familydefault}{\mddefault}{\updefault}{\color[rgb]{0,0,0}e)}%
}}}}
\put(3226,-1186){\makebox(0,0)[lb]{\smash{{\SetFigFont{12}{14.4}{\familydefault}{\mddefault}{\updefault}{\color[rgb]{0,0,0}$0$}%
}}}}
\put(3826,-1186){\makebox(0,0)[lb]{\smash{{\SetFigFont{12}{14.4}{\familydefault}{\mddefault}{\updefault}{\color[rgb]{0,0,0}$5$}%
}}}}
\put(4426,-1186){\makebox(0,0)[lb]{\smash{{\SetFigFont{12}{14.4}{\familydefault}{\mddefault}{\updefault}{\color[rgb]{0,0,0}$3$}%
}}}}
\put(5026,-1186){\makebox(0,0)[lb]{\smash{{\SetFigFont{12}{14.4}{\familydefault}{\mddefault}{\updefault}{\color[rgb]{0,0,0}$1$}%
}}}}
\put(4126,-1786){\makebox(0,0)[lb]{\smash{{\SetFigFont{12}{14.4}{\familydefault}{\mddefault}{\updefault}{\color[rgb]{0,0,0}$\bot$}%
}}}}
\end{picture}%
}\end{center}
{\bf a)} depicts an arena $\A^{\bot}$ where bold arrows represent the edges of a $0$-strategy $\sigma$,
and dashed arrows represent edges not included in $\sigma$. Further, all nodes belong to player $0$,
where the numbers inside the nodes represent the colors. {\bf b)} shows the set $S_\sigma$ of strict improvements
w.r.t. $\sigma$. {\bf c)} The heuristic applied usually for choosing a deterministic direct improvement
of $\sigma$ is to take a maximal subset of $S_{\sigma}$ so that for every node, for which a strict improvement
exists, there is exactly one strict improvement chosen. In this example this leads to the strategy depicted in c).
{\bf d)} The algorithm presented in this article, on the other hand, chooses the non-deterministic strategy $I_{\sigma} = \sigma \cup S_{\sigma}$, as shown in d). {\bf e)} Calculating the valuation of both $I_{\sigma}$, and the strategy shown in e) shows that both strategy are equivalent w.r.t. their valuation (see also lemma~\ref{lem:deterministic-optimal}). This means the strategy $I_\sigma$ is already optimal
in difference to c).

\section{Missing Proofs}

\subsection{Preliminaries}

\begin{definition}
Given an arena $\A=(V,E,o)$ and a target set $T \subseteq V$ of nodes, 
we define the $i$-attractor $\attr_i[\A](T)$ to $T$ in $\A$ by
\[
  \begin{array}{lcl}
    A_0     & := & T\\
    A_{i+1} & := & A_i \cup \{ s \in V_i | s E \cap A_i \neq \emptyset \} \cup \{ s \in V_{1-i} | s E \subseteq A_i \}\\
    \attr_0[\A](T) & := & \bigcup_{i\ge 0} A_{i}.
  \end{array}
\]
The rank $r(s) \in \N \cup \{\infty\}$ of a node $s$ w.r.t. to $\attr_0[\A](T)$ is given by
\[
  \min\{ i \in \N | s \in A_i \}
\]
where we assume that $\min \emptyset = \infty$.

A strategy $\sigma\subseteq E_i$ is then an $i$-attractor strategy to $T$, if for every $(s,t)\in \sigma$
the rank decreases along $(s,t)$ as long as $s$ has finite, non-zero rank.
\end{definition}
\begin{remark}
Obviously, player $i$ can use any $i$-attractor strategy to force any play starting from a node with finite rank
into $T$ on an acyclic path as the rank is strictly decreasing until $T$ is hit.
\end{remark}

\subsection{Parity Game Arenas with Escape for Player $0$}
\begin{qlemma}{\ref{lem:cycle-weight}}
Assume that $\chi = s_0 s_1 \ldots s_n$ is a non-empty cycle in the parity game arena $\mathcal{A}$, i.e. $s_0 \in s_n E$ and $n \ge 0$. 
$\chi$ is $0$-dominated, i.e. the highest color in $\chi$ is even if and only if $\wp(\chi) \succ \zp$.
$\chi$ is $1$-dominated if and only if $\wp(\chi) \prec \zp$.
\end{qlemma}
\begin{prf}
Wlog. we may assume that $s_0$ has the dominating color in $\chi$.
As all remaining nodes in $\chi$ have at most color $c(s_0)$, the color profile $\wp(\chi)$ is $0$
for all colors $> c(s_0)$. Hence, the highest color in which $\wp(\chi)$ and $\zp$ differ is $c(s)$.
If $c(s)$ is even, then $\wp(\chi) \succ \zp$ by definition, otherwise $\wp(\chi) \prec \zp$, as $\wp(\chi)_{c(s_0)} > 0$.
The other direction is shown similarly.
\end{prf}

\begin{qtheorem}{\ref{thm:w0-unchanged}}
Player $i$ wins the node $s$ in $\mathcal{A}$ iff he wins it in $\mathcal{A}_\bot$.
\end{qtheorem}
\begin{prf}
Let $\sigma^\ast_i$ be the optimal, memoryless winning strategy in the parity game $\A$, 
and $W_i$ the winning set of of player $i$ w.r.t. $\sigma^\ast_i$.

First consider the case $s\in W_0$. 
As only player $0$ can choose to move to $\bot$, 
any play $\pi$ in $\mathcal{A}_\bot$ w.r.t. $\sigma^\ast_0$ is a play in $\mathcal{A}$, too.
Hence, $\pi$ is infinite, and won by player $0$ w.r.t. the parity game winning condition.
Thus, $\pi$ has the value $\infty$.

Assume now that $s\in W_1$.
Player $1$ can use his optimal strategy to force player $0$ starting from $s$ 
 into a play such that every cycle visited is $1$-dominated. If player $0$ does not move to $\bot$,
 the infinite play also exists in the original parity game arena, is therefore won by player $1$,
 and, hence, has the value $-\infty$ in the escape game.
On the other hand, in the escape parity game $\mathcal{A}_\bot$ player $0$ has now
the option to escape any such infinite play by opting to terminate the game by moving to $\bot$. 
Consider therefore a finite play $\pi = s_0 s_1 \ldots s_n \bot$. 
Assume that this path is not acyclic.
Thus, as we are only counting how often a given color appears along the path, we may split $\pi$ into a simple path $\pi'$ from $s_0$
to $\top$ and several cycles $\chi_1,\ldots,\chi_l$. 
By using his winning strategy $\sigma^\ast_1$ player $1$ can make sure that every
such cycle has an odd color as maximal color. 
It is now easy to see that $\wp(\chi_j) \prec \zp$ by definition of $\prec$. 
Thus, we have
\[
	\wp(\pi) = \wp(\pi') + \wp(\chi_1) + \ldots + \wp(\chi_l) \prec \wp(\pi') \preceq \overline{\wp}.
\]
\end{prf}

\subsection{Strategy Improvement}

\begin{qlemma}{\ref{lem:bellman-ford}}
Let $\sigma\subseteq E_0$ be a reasonable strategy of player $0$.
We define $\val_{\bot}: V\cup\{\bot\} \to \mathcal{P}$ 
by $\val_{\bot}(\bot) := \zp$, and $\val_{\bot}(s) = \infty$ for all $s\in V$,
and the operator $F_\sigma : (V\cup\{\bot\} \to \mathcal{P}) \to (V\cup\{\bot\} \to \mathcal{P})$ by
\[
    \begin{array}{lcll}
	 	F_{\sigma}[\val](\bot) & := & \zp &\\
	 	F_{\sigma}[\val](s)    & := & \wp(s) + \min^{\preceq} \{ \val(t) \mid (s,t)\in E_1 \} & \text{ if } s\in V_1,\\
	 	F_{\sigma}[\val](s)    & := & \wp(s) + \max^{\preceq} \{ \val(t) \mid (s,t)\in \sigma \} & \text{ if } s\in V_0,
		\end{array}
\]
for any $\val : V\cup\{\bot\} \to \mathcal{P}$.

Then, the valuation $\vals$ of $\sigma$ is given
as the limit of the sequence $F^i_\sigma[\val_\bot]$ for $i\to \infty$, 
and this limit is reached after at most $\abs{V}$ iterations.
\end{qlemma}
\begin{prf}
For all $\val,\val' : V\cup\{\bot\} \to \mathcal{P}$ with $\val(s) \preceq \val'(s)$ for $s\in V\cup\{\bot\}$
we have $F_{\sigma}[\val](s) \preceq F_{\sigma}[\val'](s)$, too, i.e. $F_\sigma$ is monotone.
Obviously, we have $F_\sigma[\val_\bot](s) \preceq \val_{\bot}(s)$ for all $s\in V\cup\{\bot\}$.
Therefore, $F^i_{\sigma}[\val_\bot](s)$ is monotonically decreasing for $i\to\infty$.

As $\sigma$ is reasonable, $\vals(s) \succ -\infty$, and it can only be finite,
if $s$ is in the $1$-attractor to $\bot$ in $\Ab|_{\sigma}$.
Further, for $\vals(s) \prec \infty$, 
$\vals(s)$ has to be the value  of an acyclic play $\pi$ in $\Ab|_{\sigma}$.
One therefore checks easily that $\vals$ is a fixed point of $F_\sigma$;
hence, by the monotonicity of $F_\sigma$, and $\vals\preceq \val_{\bot}$,
we have $\vals \preceq F^i[\val_\bot]$ for all $i\in\N$.

Let $C_i$ be the set of nodes $s\in V\cup\{\bot\}$ such that $F_\sigma^i[\val_\bot](s) = \vals(s)$.
Obviously, we have $\bot\in C_i$ for all $i\in\N$.
As $F^i_{\sigma}[\val_\bot]$ is monotonically decreasing, and bounded from below by $\vals$,
 we have $C_i \subseteq C_{i+1}$.

Define $B_i$ to be the boundary of $C_i$, i.e. the set of nodes $s\in V\setminus C_i$ with
$sE\cap C_i \neq \emptyset \wedge sE \cap V\setminus C_i \neq \emptyset$.

If $B_i \subseteq V_0$, then player $0$ has a strategy to stay away from $\bot \in C_i$
for every node $s\in V\setminus C_i$. It is easy to see that $F^i[\val_\bot](s) = \infty$
for all $s\in V\setminus C_i$ in this case.

Thus, assume $B_i \cap V_1 \neq \emptyset$.
As player $1$ eventually needs to enter $C_i$ in order to reach $\bot$,
he has to use an edge from a node $s'\in V_1\cap B_i$ to $C_i$.
At least for this node $s'$ we have to have $s'\in C_{i+1}$.

Hence, we have to have $C_i = V$ for some $i \le \abs{V}$, implying 
$F_\sigma^{i+1}[\val_{\bot}] = F_\sigma^i[\val_\bot]$.
\end{prf}

\begin{definition}
We write $\tau_\sigma \subseteq E_1$ for the $1$-strategy consisting of the edges $(s,t)$ with 
$\vals( s ) = \wp(s) + \vals(t)$.
\end{definition}

\begin{qcorollary}{\ref{cor:reasonable-impr}}
If $\sigma$ is reasonable, then any direct improvement $\sigma'$ of $\sigma$ is reasonable, too.
\end{qcorollary}
\begin{prf}
For any cycle $s_0 s_1 \ldots s_l$ with $s_0 \in s_l E_\sigma$, we have
\[
	\vals(s_0) \preceq \wp( s_0 \ldots s_l )+\vals(s_0) \text{, i.e. } \zp \prec \wp(s_0 \ldots s_l).
\]
\end{prf}

\begin{qcorollary}{\ref{cor:impr-improves}}
Let $\sigma$ be a reasonable strategy.
\begin{itemize}
\item[(a)] For a direct improvement $\sigma'$ of $\sigma$ we have that 
$\val_{\sigma}(s) \preceq \val_{\sigma'}(s)$ for all $s\in V$.
\item[(b)] If $(s,t)\in\sigma'$ is a strict improvement of $\sigma$, 
then $\val_{\sigma}(s) \prec \val_{\sigma'}(s)$.
\end{itemize}
\end{qcorollary}
\begin{prf}
(a) Let $s$ be any node.
For any play $\pi = s_0 s_1 \ldots s_n \bot$ starting from $s$ in $\Ab|_{\sigma}$ we have 
already shown:
\[
  \val_\sigma(s) \preceq \wp(\pi) +\vals(\bot) = \wp(\pi) \preceq \val_{\sigma'}(s).
\]
(b) As $(s,t)$ is a strict improvement of $\sigma$, 
we have (i) $\val_{\sigma}(s) \prec \wp(s) + \val_{\sigma}(t)$, (ii) $s\in V_0$, and, hence,
(iii) $\val_{\sigma'}(s) = \max^{\prec}\{ \wp(s) + \val_{\sigma'}(t') \mid (s,t')\in\sigma' \}$.
With the result from (a) it follows that
\[
\val_{\sigma}(s) \prec \wp(s) + \val_{\sigma}(t) \preceq \wp(s) + \val_{\sigma'}(t) \preceq \val_{\sigma'}(s).
\]
\end{prf}

\begin{qlemma}{\ref{lem:impr-till-optimal}}
As long as there is a node $s\in W_0$ with $\val_{\sigma}(s) \prec \infty$, $\sigma$ has at least one strict improvement.
\end{qlemma}
\begin{prf}
Let $A$ be the set of nodes $t$ with $\val_{\sigma}(t) \prec \infty$, i.e. $A$ is the $1$-attractor to $\bot$
in $\Ab|_{\sigma}$. 
By assumption we have $W_0 \cap A \neq \emptyset$.
Assume $s\in A\cap W_0$. 
Let $\pi$ be any play determined by $\tau_{\sigma}$ and $\sigma^\ast_0$.
As $\sigma^\ast_0$ is optimal and $s\in W_0$, $\pi$ stays in $W_0$ forever, i.e. the play
is infinite.

First, assume $\pi$ does not leave $A$. 
Every time $\pi$ uses an edge $(u,v)$ which does not exist in $\mathcal{A}_\bot |_\sigma$ it has to hold that $u\in V_0$. 
Hence, as $\sigma$ is not strict improvable, we have to have $\val_{\sigma}(u) \succeq \wp(u) + \val_{\sigma}(v)$
for all edges $(u,v)\in \sigma^\ast_0$.
On the other hand, we have $\val_{\sigma}(u) = \wp(u) + \val_{\sigma}(v)$
along edges $(u,v)\in\tau_{\sigma}$.
Thus, the value of any cycle visited by $\pi$ is $\prec \zp$ --
a contradiction.

Therefore, consider the case that $\pi$ leaves $A$. 
This also has to happen along an edge $(u,v)$ with $u \in V_0$.
As $u\in A$ and $v\in V\setminus A$ we have $\val_{\sigma}(u) \prec \infty = \val_{\sigma}(v)$. 
Hence, $(u,v)$ is a strict improvement.
\end{prf}

\begin{qlemma}{\ref{lem:deterministic-optimal}}
Let $\sigma$ be a reasonable strategy of player $0$ in $\Ab$, \
and $\sopt$
the strategy consisting of all improvements of $\sigma$.

Then every deterministic strategy $\sigma'\subseteq \sopt$ with $\val_{\sopt}(s) = \wp(s) + \val_{\sopt}(t)$
for all $(s,t) \in \sigma'$ satisfies $\val_{\sopt} = \val_{\sigma'}$.
\end{qlemma}
\begin{prf}
By definition, $\sigma'$ is a direct improvement of $\sopt$, 
hence, we have $\val_{\sopt}(s) \preceq \val_{\sigma'}(s)$ for 
all nodes $s$.

On the other hand, $\sigma'$ is also a direct improvement of $\sigma$,
as $\sigma' \subseteq \sopt$. Thus, we have $\val_{\sigma'}(s) \preceq \val_{\sopt}(s)$
for all $s\in V$.
\end{prf}

\begin{lemma}{\label{lem:optimal-strategy}}
\begin{enumerate}
\item[(a)]
For $\sigma_a$ and $\sigma_b$ two reasonable strategies of player $0$, we define the strategy $\sigma_{ab}$
by 
\[
	(s,t) \in \sigma_{ab} :\Leftrightarrow \max\{ \val_{\sigma_a}(s), \val_{\sigma_b}(s) \} \preceq \wp(s)+ \max^\prec\{ \val_{\sigma_a}(t), \val_{\sigma_b}(t) \}.
\]
Then $\max^\prec\{ \val_{\sigma_a}(s), \val_{\sigma_b}(s) \} \preceq \val_{\sigma_{ab}}(s)$ for all $s\in V$, i.e.
 there is a strategy $\hat{\sigma}$ such that for all other strategies $\sigma$ we have $\val_{\sigma}(s) \preceq \val_{\hat{\sigma}}(s)$ for all $s\in V$.
\item[(b)] If $\val_{\sigma}(s) \prec \val_{\hat{\sigma}}(s)$ for at least one $s\in V$, then $\sigma$ has a strict improvement.
\end{enumerate}
\end{lemma}
\begin{prf}
(a) We first show that $\sigma_{ab}$ is indeed a strategy. 
Consider any $s\in V_0$. Then there is at least one $t_a$ s.t. $(s,t_a)\in \sigma_a$ and $\val_{\sigma_a}(s) = \wp(s) + \val_{\sigma_a}(t_a)$,
and similarly a $t_b$ with the same properties w.r.t. $\sigma_b$.
Assume $\val_{\sigma_a}(s) \preceq \val_{\sigma_b}(s)$ -- the other case being similar.
By definition of $\vals$ we then have
\[
	\val_{\sigma_a}(t_b) \preceq \val_{\sigma_a}(t_a) = \wp(s)+\val_{\sigma_a}(s) \preceq \wp(s) + \val_{\sigma_b}(s) = \val_{\sigma_b}(t_b),
\]
i.e. $(s,t_b)\in \sigma_{ab}$.

By definition, we have 
\[
 \max\{ \val_{\sigma_a}(s), \val_{\sigma_b}(s) \} \preceq \wp(s)+ \max^\prec\{ \val_{\sigma_a}(s), \val_{\sigma_b}(s) \} \quad (*)
\]
along every edge $(s,t)\in \sigma_{ab}$. 
For any edge $(s,t) \in E_1$, we have
\[
	\val_{\sigma_a}(s) \preceq \wp(s) + \val_{\sigma_a}(t) \text{ and } \val_{\sigma_b}(s) \preceq \wp(s) + \val_{\sigma_b}(t).
\]
Hence, $(*)$ holds along every edge of $\mathcal{A}_\bot |_{\sigma_{ab}}$. Therefore, any cycle in $\mathcal{A}_\bot |_{\sigma_{ab}}$
has to be $0$-dominated, again, i.e. $\sigma_{ab}$ is reasonable, too.

If $\val_{\sigma_{ab}}(s) = \infty$, there is nothing to show.
Assume $\val_{\sigma_{ab}}(s) \prec \infty$, first, and let $\pi = s_0 s_1 \ldots s_n \bot$ be any acyclic play with
$\wp(\pi) = \val_{\sigma_{ab}}(s)$. Because of $(*)$ we then have $\max\{ \val_{\sigma_a}(s), \val_{\sigma_b}(s) \} \preceq \wp(\pi) = \val_{\sigma_{ab}}(s)$, again.

(b) If there is some node $s\in V$ with $\val_{\sigma}(s) \prec \val_{\hat{\sigma}}(s) = \infty$, we already know that
$\sigma$ has a strict improvement as it is not optimal ($s$ is won by $\hat{\sigma}$ but not by $\sigma$).

Therefore assume that $\val_{\hat{\sigma}}(s') =\infty$ implies $\vals(s') = \infty$ for all nodes $s'$,
and let $s$ be a node with $\val_{\hat{\sigma}}(s) \prec \infty$. 
Let $\pi$ again be an acyclic
play in $\mathcal{A}_\bot |_{\hat{\sigma},\tau_{\sigma}}$ with $\wp(\pi) = \val_{\hat{\sigma}}(s)$,
i.e. player $0$ uses $\hat{\sigma}$ and player $1$ his response-strategy $\tau_{\sigma}$ for $\sigma$.

As $\sigma$ has no strict improvements, we have $\val_{\sigma}(s) \succeq \wp(s) + \val_{\sigma}(t)$ 
for all edges $(s,t) \in E_0$; on the other hand, along the edges $(s,t)\in\tau_{\sigma}$ we have 
$\val_{\sigma}(s) = \wp(s) + \val_{\sigma}(t)$ by definition of $\tau_\sigma$.

Hence, we get $\val_{\hat{\sigma}}(s) \succeq \val_{\sigma}(s) \succeq \wp(\pi) = \val_{\hat{\sigma}}(s)$,
if $\sigma$ has no strict improvements.
\end{prf}

\begin{qproposition}{\ref{prop:dij}}
$\val_{I_{\sigma}}$ can be calculated using Dijkstra's algorithm which needs $O(\abs{V}^2)$
operations on color-profiles on dense graphs;
for graphs whose out-degree is bound by some $b$ this can be improved to $O(b\cdot \abs{V} \cdot \log \abs{V})$ by
using a heap. 
\end{qproposition}
\begin{prf}
Let $\sigma$ be a reasonable $\sigma$ strategy of player $0$,
and $A$ the $1$-attractor to $\bot$ in $\Ab|_{\sopt}$.
For all nodes $s\in V\setminus A$, we have $\val_{\sopt}(s) = \infty$.
We therefore have only to consider the graph $(A, E_{\sopt}\cap A\times A)$ 
in order to calculate $\val_{\sopt}$ for the nodes in $A$.

Recall that we have for every edge $(u,v)$ in $\Ab|_{\sopt}$ 
that $\val_{\sigma}(u) \preceq \wp(u) + \val_{\sigma}(v)$.
Define now for $(u,v) \in E_{\sopt}\cap A\times A$ the function $w$ by 
$w(u,v) := (\wp(u)+ \vals(v)) - \vals(u) \succeq \zp$.
Hence, for any path $\pi' = t_0 t_1 \ldots t_n \bot$ in $(A,E_{\sopt}\cap A\times A)$ we have
\[
\begin{array}{cl}
	& \wp(\pi') - \val_{\sigma}(t_0)\\
 = & \wp(t_0) + \ldots + \wp(t_n) + ( \vals(t_1) - \vals(t_1) ) + \ldots + (\vals(t_n) - \vals(t_n)) - \vals(t_0)\\
 = & ( \wp(t_0)+ \val_{\sigma}(t_1) - \val_{\sigma}(t_0)) + \ldots + ( \wp(t_n) + \val_{\sigma}(\bot) - \val_{\sigma}(t_n))\\
 = & w(t_0,t_1) + w(t_1,t_2) + \ldots + w(t_n,\bot).
\end{array}
\]
Therefore, for any $s\in A$ we have that $\val_{\sopt}(s)$, i.e.
the $\prec$-minimal value player $1$ can guarantee to achieve in a play starting from $s$,
has to be $\vals(s)$ plus the $\prec$-minimal value $\delta_\sigma(s)$ player $1$ can guarantee starting from $s$
in the edge-weighted graph $(A,E_{\sopt}\cap A\times A,w)$.

As $w(u,v) \succeq \zp$, 
we can use Dijkstra's algorithm to find $\delta_{\sigma}(s)$ 
with the restriction that we only may add a node controlled by player $0$ to the boundary in every 
step of Dijkstra's algorithm, if all successors of this node have already been evaluated.
We then have $\delta_{\sigma}(s) = \val_{\sopt}(s) - \val_{\sigma}(s)$.
\end{prf}

\subsection{Comparison with the Algorithm by Jurdzinski and V{\"o}ge}

\begin{qtheorem}{\ref{thm:steps}}
Let $\Ab$ be a escape-parity-game arena where every node of player $0$ has at most
two successor. Then the number of improvement steps needed to reach an optimal winning
strategy is bound by
$3\cdot 1.724^{\abs{V_0}}.$
\end{qtheorem}
\begin{prf}
\begin{ass}
We assume that player $0$ can only choose between at most two different successors
in every state controlled by him, i.e. $\forall v\in V_0: \abs{vE} \in \{1,2\}$.
\end{ass}

Let $(\sigma_\bot = \sigma_0) \prec \sigma_1 \prec \ldots \prec (\sigma_l = \hat{\sigma})$
be the sequence of strategies produced by the strategy-improvement algorithm presented in this article.
As already shown, we may assume that $\sigma_i$ is deterministic.

For $\sigma_i$ let $k_i$ be the number of nodes $s\in V_0$ such that there
is at least one strict improvement of $\sigma$ at $s$, i.e.
\[
	k_i := \abs{\src(S_{\sigma_i})} \text{ with } \src(S_{\sigma_i}) := \{ s\in V_0 \mid \exists (s,t) \in S_{\sigma_i} \}.
\]
(Recall that $S_\sigma$ is defined to be the set of strict improvements of a given strategy $\sigma$.)

Then there are at least $2^{k_i}-1$ deterministic direct improvements $\sigma'$ of $\sigma_i$ with
$\sigma_i \prec \sigma'$ and $\sigma'\setminus \sigma_i \subseteq S_{\sigma_i}$.
\footnote{Note that we do not claim that $\sigma_{i+1}$ is one of these strategies $\sigma'$.}

We then have $\sigma_i \prec \sigma' \preceq \sigma_{i+1}$ for every such $\sigma'$.
Now, as $\sigma_i \prec \sigma_{i+1}$, we know that every such $\sigma'$ 
has not been considered in a previous step ($<i$) nor will it be considered
in any following step ($>i$).
Therefore, at least $2^{k_i} - 1$ new deterministic strategies can be ruled out as candidates for
optimal winning strategies.

Hence, if  $S_k$ is the number of deterministic strategies which have at most $k$
nodes at which there exists at least one strict improvement, 
we get as an upper bound for the number of improvement steps
\[
	S_k + \frac{2^{\abs{V_0}}}{2^{k+1}-1} \le S_k + 2^{\abs{V_0}-k}.
\]
The next lemma bounds the number $S_{k_i}$ of strategies $\sigma_i$ having the same value for $k_i$:
\begin{lemma}\label{lem:bound-on-strategies}
Let $(\sigma_i)_{0\le i \le l} = \sigma_\bot = \sigma_0 \prec \sigma_1 \prec \ldots \prec \sigma_l = \hat{\sigma}$
be the sequence of reasonable deterministic strategies generated by the strategy improvement algorithm.

For an arena $\Ab$ with $\abs{sE} \le 2$ for all $s\in V_0$ it holds that
 there are  most ${\abs{V_0} \choose k'}$ strategies in $(\sigma_i)_{0\le i \le l}$ with $\abs{\src(S_{\sigma_i})} = k'$.
\end{lemma}
\begin{prf}
First note the following easy fact: 
As along any edge $(s,t)\in \sigma$ holds, we have $\vals(s) \succeq \wp(s) + \vals(t)$ by definition of $F_{\sigma}$.
Thus, for any strategy $\sigma\subseteq E_0$ of player $0$ it holds that $S_{\sigma} \cap \sigma = \emptyset$.

Next, let $\sigma_a$ and $\sigma_b$ be two reasonable strategies of player $0$ in $\Ab$.
We claim that it holds that 
\begin{enumerate}
\item[(a)] 
  If $S_{\sigma_b}\cap \sigma_a = \emptyset$,
  %i.e. $\sigma_a$ uses no edge which is a strict improvement of $\sigma_b$,
  we have $\sigma_a \preceq \sigma_b$.
\item[(b)] 
  %For any set $K$ of edges set $\src(K) := \{ s \in V \mid \exists t\in V : (s,t) \in K \}$.
  Assume that $\abs{sE} \le 2$ for all $s\in V_0$.
  If $\src(S_{\sigma_b}) \subseteq \src(S_{\sigma_a})$, it holds that $\sigma_a \preceq \sigma_b$.
\end{enumerate}
Before given the proofs to these two claims, note that (b) already implies 
that we can have at most ${ \abs{V_0} \choose k'}$-many strategies $\sigma_i$ with $k_i = k'$,
as this is the number of disjoint subsets of $V_0$ with $k'$ distinct elements.

In order to show (b), we first need to show (a):
(a) Let $\Ab'$ be the arena resulting from $\Ab$ by removing all strict improvements of $\sigma_b$ from $E$, i.e.
$E' = E \setminus S_{\sigma_b}$. %E_1 \cup \{ (s,t) \in E_0 \mid \val_{\sigma}(s) \succeq \wp(s) + \val_{\sigma}(t)\}$.
Both $\sigma_a$ and $\sigma_b$ are reasonable strategies of player $0$ in $\Ab'$, 
as we only remove edges and these edges are neither used by $\sigma$ nor by $\sigma'$.
This also means that
the operators $F_\sigma$ and $F_{\sigma'}$ stay unchanged,
implying that the valuations of $\sigma$ (reps. $\sigma'$) on $\Ab$ and $\Ab'$ coincide.
But as $\sigma_b$ has no strict improvements in $\Ab'$, it has to hold that $\sigma_b$ is an optimal winning strategy
in $\Ab'$, meaning that $\sigma_a \preceq \sigma_b$ (cf. lemma~\ref{lem:optimal-strategy}).

(b) 
Set $C = S_{\sigma_b} \cap \sigma_a$.
For every $s\in\src(C)$ we find a $t_C$ such that $(s,t^C_s) \in C$, 
a $t^{\sigma_b}_{s}$ with $(s,t^{\sigma_b}_s) \in \sigma_b$ (as $\sigma_b$ is a strategy),
and a $t^{S_{\sigma_a}}_s$ with $(s,t^{S_{\sigma_a}}_s)\in S_{\sigma_a}$ (as $\src(S_{\sigma_b}) \subseteq \src(S_{\sigma_a})$).

Now, because of $S_{\sigma} \cap \sigma = \emptyset$ for any strategy $\sigma$,
we may conclude that $t^C \neq t^{\sigma_b}_s$, and $t^C\neq t^{S_{\sigma_a}}_s$ for all $s\in \src(C)$.
Thus, as we assume that $\abs{sE} \le 2$, it has to hold that $t^{S_{\sigma_a}}_s = t^{\sigma_b}_s$ for all $s\in\src(C)$.
We define therefore $C' = \{ (s,t^{\sigma_b}_s) \mid s\in\src(C) \}$, and
\[
	\sigma' := C' \cup \sigma_a\setminus C.
\]
As $C'\subseteq S_{\sigma_a}$, we have $\sigma_a \preceq \sigma'$.
Further $\sigma' \preceq \sigma_b$, as $\sigma' \cap S_{\sigma_b} = \emptyset$.
\end{prf}
The last lemma can be found in \cite{MS99} for Markov decision processes.

As long as $1 \le k \le \frac{\abs{V_0}}{3}$, we have
\[
S_k \le \sum_{k' = 0}^{k} { \abs{V_0} \choose k' } \le 2 { \abs{V_0} \choose k } \le 
2 
\left( \frac{\abs{V_0}}{k} \cdot e \right)^k.
\]
What remains is to find a $1 \le k \le \frac{\abs{V_0}}{3}$ such that
\[
2
 \left( \frac{\abs{V_0}}{k} \cdot e \right)^k + 2^{\abs{V_0} - k}
\]
is minimal. 
For this set $b = \frac{\abs{V_0}}{k}$ with $b\ge 3$, yielding
\[
2 
\cdot e^{\abs{V_0} \cdot \frac{1+\ln b}{b}} + e^{\ln 2 \cdot \abs{V_0} \cdot \frac{b-1}{b}}.
\]
As $\frac{1+\ln b}{b}$ is strictly decreasing and $\frac{b-1}{b}$ is strictly increasing, we need to look
for the largest $b\ge 3$ such that
\[
	\frac{1+\ln b}{b} \ge \ln2 \cdot \frac{b-1}{b}.
\]
Using e.g. Newton's method one can easily check that $b \in (4.6,4.7)$ with $b\approx 4.66438$.
We therefore get
\[
 3\cdot e^{0.545\cdot \abs{V_0}} 
 \le 
 3\cdot 1.724^{\abs{V_0}} 
 \le 
 3\cdot 1.313^{\abs{V}}
\]
as an alternative upper bound for the number of improvement steps for an arena with out-degree two
\footnote{Using a more detailed analysis in the spirit of \cite{BSV02} one can even show an upper bound
of $O( 1.71^{\abs{V_0}})$.}.

\end{prf}

\end{document}